# Standalone micro-reformer for on-board hydrogen production from dimethyl ether

*M. Bianchini[1], N. Alayo[1], L. Soler[2], M. Salleras[3], L. Fonseca[3], J. Llorca[2], A. Tarancón[1,4]*

1. Catalonia Institute for Energy Research (IREC), Department of Advanced Materials for Energy, 08930, Sant Adriá del Besòs, Barcelona, Spain
2. Institute of Energy Technologies, Department of Chemical Engineering and Barcelona Research Center in Multiscale Science and Engineering, Universitat Politècnica de Catalunya, EEBE, Eduard Maristany 10-14, 08019 Barcelona, Spain
3. IMB-CNM (CSIC), Institute of Microelectronics of Barcelona, National Center of Microelectronics, CSIC, Campus UAB, 08193, Bellaterra, Barcelona, Spain
4. ICREA, 08010, Barcelona, Spain

## Abstract

Entering a new era of sustainable energy generation and consumption, new solutions for powering consumer electronics are required to tackle the limited capacity provided by the portable power sources employed nowadays. Hydrocarbon-fed micro-fuel cells represent a promising technology for this purpose, and micro-reactor technology can indeed enable their integration for portable applications. In this work, we present the design and fully scalable wafer-level fabrication of a MEMS-based catalytic micro-reactor, paving the way towards on-board hydrogen production for portable power generators. The device consists of an array of thousands of vertically-aligned micro-channels, 500 μm in length and 50 μm in diameter, for an overall superficial area per unit volume of 120 $cm^2$ $cm^{-3}$ and it embeds a thin-film heater for efficient reaction start-up. Functionalization of the active area was achieved by atomic layer deposition, resulting in the uniform coating of a $Pt/Al_2O_3$ heterogeneous catalyst. The temperature-dependent dimethyl ether (DME)-to-syngas conversion is tested through steam reforming (SR) and partial oxidation (POX) reactions. Here, conversion rates up to 74% and hydrogen selectivity of ~60% are obtained by steam reforming at 650 ºC, while a specific volumetric hydrogen production of 4.5 $mL_{H2}$ $mL^{-1}_{DME}$ $cm^{-3}_{REACTOR}$ at 600 ºC is obtained from DME POX in a standalone device tested by means of a 3D printed ceramic housing.

## Keywords

MEMS; micro-reactor; Dimethyl ether; micro-power sources; Partial oxidation; steam reforming

## 1. Introduction

Since the beginning of this century, technological advancements in microelectronics made it possible to reduce the size of portable devices while achieving greater quality and efficiency. However, the power consumption rises proportionately to the new functionalities these devices are able to operate[1]. The last generation of portable devices mostly relies on lithium-ion batteries, in spite of the limited energy density they provide. In order to fulfill the energy gap between power consumption and the specific energy provided by portable power generators, innovative approaches to generate electric power from fuels have been explored[2–8].

Entering a new era of sustainable energy generation and consumption based on renewable energies, hydrogen is expected to tackle an important role as green energy carrier with the greatest gravimetric energy density and low contamination[9]. Nevertheless, it suffers from many drawbacks related with safety and storage when it comes to portable applications. The use of catalytic reformers for hydrogen and synthesis gas (syngas) production is an established alternative to direct hydrogen storage since hydrocarbon-based fuels also provide high energy density, low cost, safety, and easy transportation[10].

Therefore, micro-scale catalytic reactors have been pointed out as a solution for the integration of fuel-based micro-power sources in portable devices. Indeed, they present many advantages compared to its macro-scale counterparts: higher surface-to-volume ratio and shorter travel distance in the reactor favor the overall reaction kinetics as heat and mass transfer are greatly enhanced. Moreover, a lower amount of reactants together with faster and more efficient mixing allow for safer operating conditions, especially suitable for strong exothermic reactions at harsh conditions[11–13].

While a multitude of micro-reactors fabrication techniques have been reviewed for a wide range of construction materials[14], combination with advances in integrated circuit (IC) and micro-electromechanical systems (MEMS) arouse great interest in the scientific community, enabling the fabrication of high-aspect-ratio structures with high accuracy and allowing for the integration of functional features such as temperature sensors and micro-heaters onto the same platform[15]. These techniques (photolithography, wet and dry etching, chemical and physical vapor depositions) can be effectively used to create micro-channels on polymer, glass or silicon substrates. Indeed, silicon is an ideal material for the fabrication of micro-reactors because of its large operating temperature range and chemical inertness.

MEMS-based micro-reactors, first demonstrated between 1987 and 1994 by DuPont scientists, can be classified into three main types[16]: i) packed bed, ii) wall coated and iii) micro-structured reactors. Each of this reactor design[12,15,17,18] presents some advantages and drawbacks depending on the chosen application. Nevertheless, certain requirements must be taken into account for continuous operation at the high temperature regime employed during hydrocarbons conversion reactions. For instance, packed-bed micro-reactors and long in-plane micro-channels reactors usually require large area and show poor heat transfer capabilities due to the insufficient radial mixing of the fuel[19]. Moreover, high-pressure drops can occur along the reaction pathways at the typical low flow rates employed at the micro-scale[20] and the catalyst stability is often compromised.

For this reason, a new design based on silicon monolith embedding an array of micro-channels was developed by Llorca and coworkers.[21,22] at multiple size scales and for the conversion of different hydrocarbons to syngas [23]. In addition, we previously demonstrated the fabrication of a MEMS-based micro-reformer consisting of an array of vertical micro-channels (50 μm diameter) defined into a silicon substrate and a serpentine-shaped heater by means of mainstream microelectronics fabrication processes. This innovative approach would allow for optimized thermal management features leading to fast start-up time (less than a minute with vacuum insulation) and low energy consumption below 60 J[24]. The device was tested for ethanol steam reforming by coating the micro-channels with a $CeO_2$ active support film and Pd/Rh nanoparticles grafted on top. Nevertheless, the use of liquid ethanol for on-board hydrogen production premises the addition of a micro-vaporizer unit, adding even more complexity to the final system.

In this work, we demonstrate the fabrication of a MEMS-based suspended micro reforming unit and its catalytic performance for conversion of dimethyl ether (DME) to syngas, especially designed for start-up of the μ-SOFC stack power generator described in brief elsewhere[25]. DME was chosen as non-petroleum based alternative fuel gas with high hydrogen-to-carbon ratio and, accordingly, high energy density. Besides, DME is non-toxic, easy to handle and its storage and transport infrastructures are the same as LPG or natural gas, highlighting its potential for worldwide adoption[26–28].

Finally, a fully scalable catalyst fabrication route based on atomic layer deposition (ALD) has been implemented, tackling one of the most commonly reported issues for coated-wall micro-reactors, i.e. the uniform surface coating and dispersion of metal nanoparticles on high-aspect-ratio structures[12]. This technique has received increasing interest in recent years for its self-limiting reaction character, leading to layer-by-layer growth of conformal coatings and outstanding precision on thickness control. This technique and its numerous applications for energy devices and catalysis have been extensively reviewed[29–32], remarking its great potential for the fabrication of homogeneous coatings or well-dispersed particles onto high-aspect-ratio structures or highly porous materials. Conversely, the most typical coating technique, i.e. dip coating (or washcoating) from a liquid suspension or sol mixture, suffers from non-uniform surface tension of the micro-reactor's walls and the resulting coating tends to collect at the corners of the reactor[12]. Here, a heterogeneous catalyst based on Pt nanoparticles dispersed onto $\gamma$-$Al_2O_3$ support was chosen as a model catalyst suitable for DME reforming[33–37]. In fact, Pt has been shown to limit coking, so that more active surface area is available for reaction[38] while $\gamma$-$Al_2O_3$ is a well-known support for its high surface area and acidic properties[39].

## 2. Experimental methods

### 2.1 Design and microfabrication

The micro-reformer unit presented in this work is designed as a monolithic reactor with an array of vertically-aligned micro-channels (50 μm diameter) etched through a silicon substrate. A serpentine-like heater is defined on top of the device for efficient start-up and the active area is thermally insulated by means of a $Si_3N_4$ membrane, preventing lateral heat losses. More details about the design and a FEM analysis are reported elsewhere[24], while a schematic view of the reactor is included in figure 1.

For the fabrication of the device, single crystal (100)-oriented p-type silicon wafers of 4 inches in diameter and 500 μm in thickness are first passivated on both sides with a dielectric bilayer consisting of 100 nm-thick thermally-grown $SiO_2$ and 300 nm of plasma-enhanced chemical vapor deposition (PECVD) $Si_3N_4$ deposited on top. Photolithography is used throughout the whole fabrication to transfer patterns onto the substrate. Three layers of 30 nm TiW, 250 nm W and 20 nm Au are sputtered on the top side of the wafer and subsequently the micro-heater is defined by lift-off of the patterned photoresist. A 2.5 μm-thick $SiO_2$ layer is deposited by PECVD on both sides of the wafer and etched by Deep Reactive Ion Etching (DRIE) on the back side of the wafer creating a mask for the micro-channels and the trench. About 480 µm of silicon are etched from the back side, and the micro-channels are eventually opened-through by repeating the same process on the top side. Finally, 1.5 µm of $SiO_2$ left on top of the heater are removed by dipping the wafers in HF 1:10 aqueous solution.

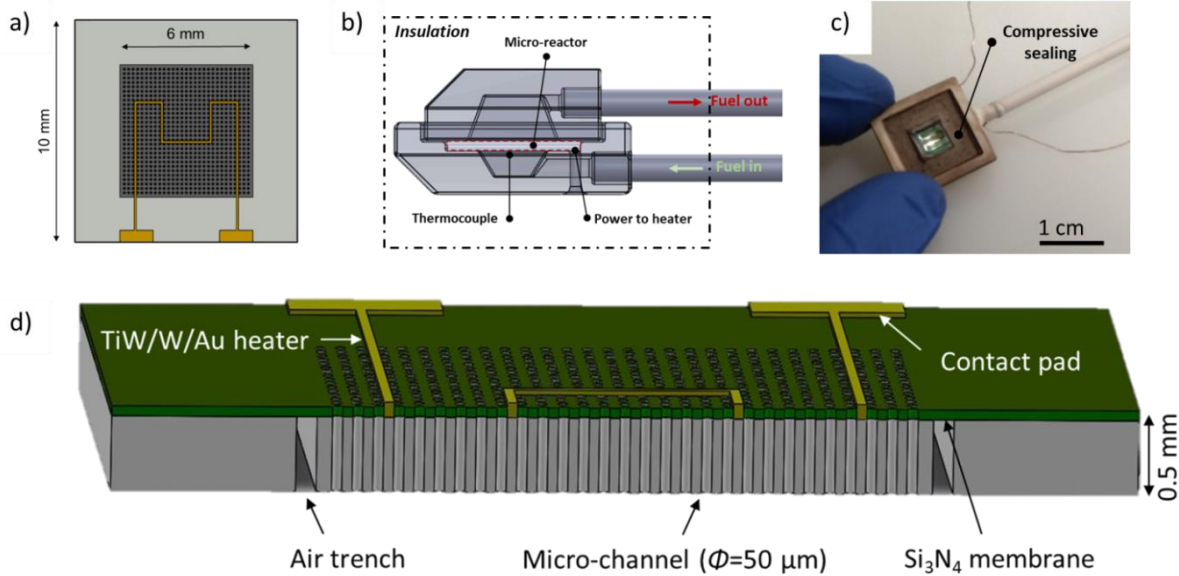

*Figure 1. a) Schematic top view of the micro-reactor; b) Schematic design of the 3D printed ceramic reactor for measurement of the standalone device; c) optical image of the ceramic housing and compressive sealing used in the experiment and d) schematic cross-section of the device with the description of its main components.*

Catalyst deposition was performed by atomic layer deposition (ALD) with a PicoSun R-200 system. Trymethylaluminium (III), 97% purity, was purchased from Sigma-Aldrich and employed as aluminum source while $H_2O$ was used as co-reactant. Each ALD cycle was composed by 0.1 sec pulse of precursor and 6 sec purging with inert gas ($N_2$). Growth-per-cycle was calibrated to be 0.83 Å/cycle and the overall processes consisted of 1200 cycles to obtained 100 nm-thick $Al_2O_3$ films. (Trimethyl)methylcyclopentadienylplatinum(IV), 99% purity, was purchased from Strem Chemicals and $O_2$ was used as co-reactant. Here, 0.8 sec pulse of platinum precursor where alternated to 1 sec pulse of $O_2$ with 12 sec of $N_2$ purge in between. 80 ALD cycles were carried out in order to obtain an array of nanoparticles grafted onto the alumina support. The temperature of the chamber was set at 300 ºC in both cases.

Crystallization of the as-deposited amorphous $Al_2O_3$ layer was obtained by rapid thermal processing (RTP) with an Annealsys As-one 100 system at a heating ramp rate of 15 ºC sec$^{-1}$ and 90 seconds dwell at the temperature set point. This fast annealing process has indeed the advantage of triggering crystallization in thin-films without damaging a MEMS device itself or the features it includes, such as the micro-heater described in this work.

## 2.2. Catalyst characterization and gas reforming

The microstructural characterization of the catalyst was performed by scanning electronic microscopy (SEM) and grazing incidence angle X-ray diffraction (GIXRD). SEM images were recorded at 5 kV using a Zeiss Neon40 Crossbeam Station instrument equipped with a field emission electron source. A backscattered electrons detector was used to observe the dispersion of Pt onto the alumina support evidenced by Z contrast. GIXRD characterization was performed with a X'Pert Pro MRD-Panalytical instrument optimizing the incident angle to 0.3° and analyzing 100 nm-thick $Al_2O_3$ layers deposited on Si (100) 1x1 $cm^2$ chips, while a Bragg-Brentano geometry was adopted to analyze the same samples after Pt deposition after ALD.

Two different catalytic reactions were carried out in order to estimate the viability of this technology for on-board hydrogen production from DME, namely steam reforming (SR) and partial oxidation (POX).

The former is the most common reaction employed for industrial purposes due to its high $H_2$ output and it occurs according to the following consecutive reactions:

$$CH_3OCH_3 + H_2O \rightleftarrows 2CH_3OH \qquad (1)$$
$$\Delta H° = 135 \text{ kJ mol}^{-1} \text{ (endothermic)}$$
$$2CH_3OH + 2H_2O \rightleftarrows 2CO_2 + 6H_2 \qquad (2)$$

Nevertheless, for a portable solid oxide fuel cell, POX is more favorable from an energy balance perspective for its exothermic character at the expenses of the $H_2$ yield and it is described by the following reaction:

$$CH_3OCH_3 + 0.5O_2 \rightleftarrows 2CO + 3H_2 \qquad \Delta H° = -37 \text{ kJ mol}^{-1} \text{ (exothermic)} \qquad (3)$$

For the evaluation of the temperature effects on the catalytic performance, a customized fixed bed reactor made of stainless steel was employed and placed inside a furnace at a temperature range between 400 °C and 650 °C (ramp rate of 3 °C $min^{-1}$). DME flow was set at 2 mL $min^{-1}$ and mixed with steam (steam to carbon ratio S/C=1.5) or air ($O_2$/DME=0.5) at the reactor inlet together with $N_2$ inert gas, accounting for a total volumetric flow of 15 mL $min^{-1}$. Gas flows were controlled with mass flows from Bronkhorst.

A customized 3D printed holder for the standalone reactor measurement (see figure 1.b) was fabricated using a Ceramaker SLA 3D printer from 3DCeram, and a gas-tight joint was implemented by compressive sealing (see figure 1.c) with ceramic felt (Fuelcellmaterials). The holder was further insulated by placing it in a glass wool bed. The presented design allows non-trivial measurements due to the following requirements: (i) the reactor must be gas-tight, (ii) insulation is needed to prevent heat losses and (iii) fluidic and electrical connection must be compatible with the temperature reached during the catalytic reactions. A fixed power of 5 W was applied to the heater and the temperature was measured with a thermocouple (type *k*) in proximity of the heater and on the frame. The heater power required for the standalone device measurement is inversely related to the level of thermal insulation of the holder. Since the holder fabricated here is made of highly conducting alumina, the power is too high. The final portable SOFC system will require much less power to reach the operating temperature according to previous simulations[25] on a system operating with ethanol and at higher operating temperature (i.e. 700 °C).

The effluent of the reactor was monitored online with an Agilent 3000A micro-GC equipped with PLOT U, Stabilwax and 5 Å Molsieve columns for a complete analysis of products.

DME conversion (%) is defined as

$$\chi_{DME} = \frac{100 \times n_{DME,conv}}{2n_{DME,in}}$$

where $n_{DME,conv}$ represents the moles of DME converted measured as the sum of moles of the carbonaceous species ($CO_2$, CO, $CH_4$ and $CH_3OH$) at the reactor outlet and $n_{DME,in}$ represents the moles of DME at the reactor inlet. Selectivity to species i (%) is calculated as the moles of i divided by the total moles of products ($H_2$, $CO_2$, CO, $CH_4$ and $CH_3OH$),

$$S_i = \frac{100 \times n_i}{\sum n_T}$$

Yield to species i (%) is calculated as

$$Y_i = \frac{\chi_{DME} \times S_i}{100}$$

All the samples where reduced under 5% $H_2$/Ar atmosphere at 300 ºC for 1h prior to the experiments to ensure that no platinum oxide was left in the catalyst. The results hereby reported were always within 5% experimental error.

Finally, the molar flows ($\dot{n}_i$) of $H_2$ ($\Delta G^\circ$ = 237 kJ mol$^{-1}$), CO ($\Delta G^\circ$ = 205.5 kJ mol$^{-1}$) and $CH_4$ ($\Delta G^\circ$ = 140 kJ mol$^{-1}$) species obtained by these reactions were used to estimate the power obtained by an ideal µ-SOFC device (assuming 100% conversion efficiency) fed with the abovementioned syngas mixture according to equation 4:

$$W = \Delta G^\circ_{H2} \cdot \dot{n}_{H2} + \Delta G^\circ_{CH4} \cdot \dot{n}_{CH4} + \Delta G^\circ_{CO} \cdot \dot{n}_{CO} \qquad (4)$$

## 3. Results and discussion

### 3.1 Micro-reactor fabrication and catalyst coating characterization

All the 4-inches processed wafers comprise more than 50 devices (figure 2.a) successfully fabricated thanks to an optimal thermalization of the specimen during the DRIE process and moreover multiple wafers were processed simultaneously, highlighting the potential of micro-fabrication approach to industrial large-scale production.

The 1x1x0.05 cm$^3$ device (figure 2.b) consisted of about 8000 micro-channels (50 µm diameter, figure 2.c) corresponding to an active area within the channels greater than 6 cm$^2$. The 50 µm-wide trench was defined on the back side, ensuring thermal insulation of the reactor frame from the catalytically active region. The surface of the heater was inspected after the HF treatment and it was found free of $SiO_2$, showing a nominal resistance R = 80 ± 2 Ω, while no macro-defects were detected on the metallic layers.

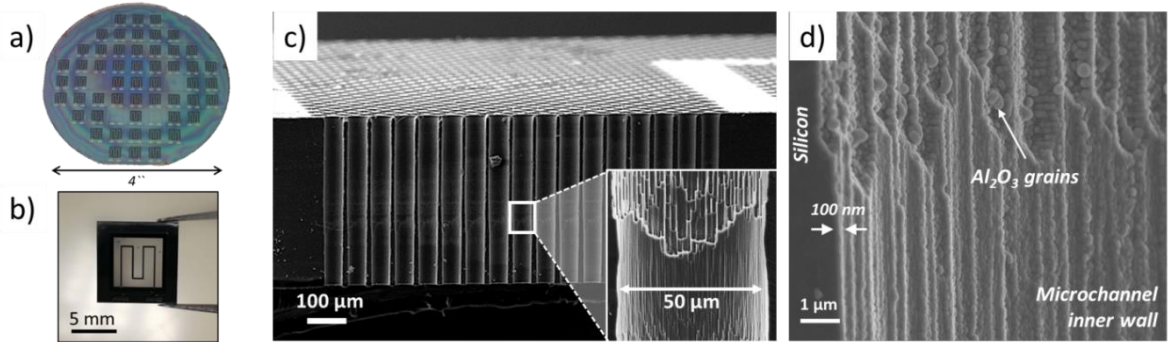

*Figure 2. a) optical image of the Si wafer after processing, comprising more than 50 devices; b) optical image of the fabricated micro-reactor showing the transparent active region where the opened micro-channels are located; c) Cross-section SEM image of the device showing the through-silicon channels and the heater path, and higher magnification of a single channel (inlet); d) Micro-channel's wall after $Al_2O_3$ deposition by ALD*

The catalyst selected in this work was fabricated by self-limiting chemical vapor deposition method, which guarantees the uniform coating of high-aspect-ratio structures such as the micro-channels presented in this work as well as the integration in the micro-fabrication process line. The 100 nm-thick $Al_2O_3$ active support layer that coated homogeneously all the micro-channels´ walls (figure 2.b) was found to be amorphous after deposition, as expected. Therefore, RTP was performed at different temperature set points and the resulting crystallographic properties examined by GIXRD (figure 3). Here, part of the spectrum (2θ = 50-60º) was not included due to the presence of Si (113) substrate reflection.

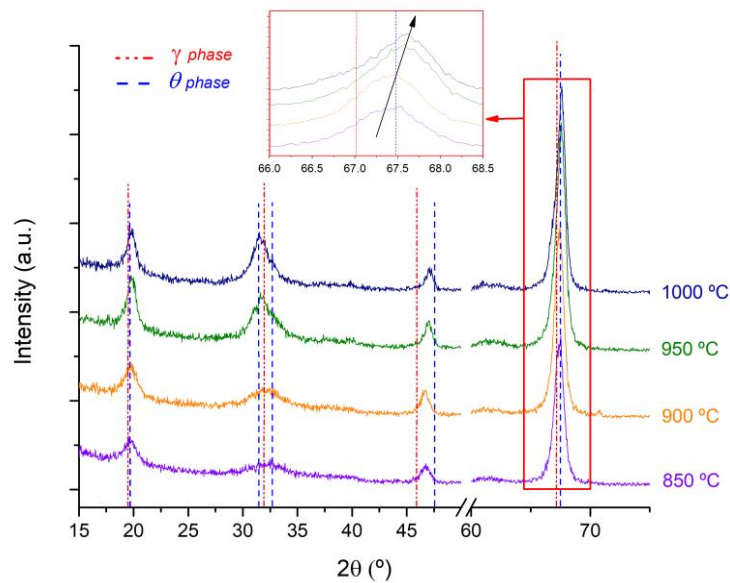

*Figure 3. GIXRD spectra of 100 nm-thick Al2O3 layers on Si(100) substrate after RTP at different temperature set points*

All the samples presented diffraction peaks related to either gamma- (JCPDS 00-020-0425) or theta- (JCPDS 00-035-0121) alumina phases. Increasing the RTP temperature set point boosts the amount of crystalline phase in the sample, as it can be observed from the evolution of the peaks around 32°, where increasing intensity of the reflection theta (004) with respect to the gamma (220). Also, figure 3 reports a shift of the most intense peak around 2θ = 67º, corresponding to the (215) theta plane and (440) gamma plane, supporting the evolving trend towards the θ- $Al_2O_3$ phase.

In this case, limiting the amount of θ-$Al_2O_3$ phase is beneficial because of its basicity, making it catalytically inactive. Conversely, γ- $Al_2O_3$ is a very common catalytic support for its acidity that is known to trigger dehydration [39,40]. Hence, a RTP temperature set point $T_{RTP}$ = 850 ºC was chosen for the fabrication of the γ- phase-rich catalytic alumina support.

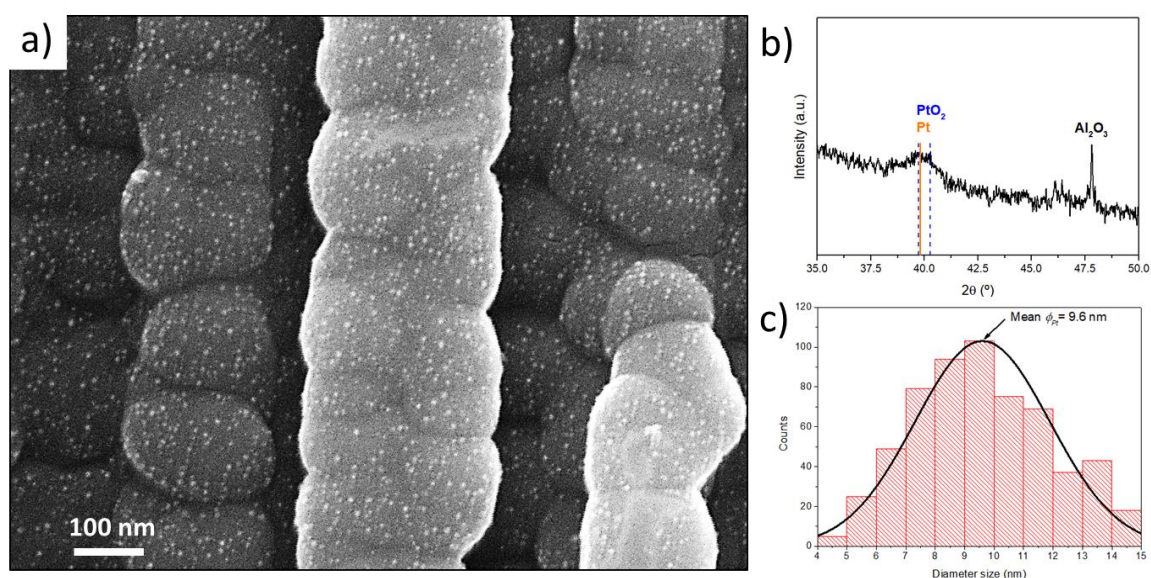

*Figure 4. a) High-magnification back-scattered electrons SEM image of the fabricated catalyst (micro-channel cross section) comprising Pt nanoparticle dispersion on $Al_2O_3$ layer; b) XRD spectrum of the catalyst deposited on Si(100) recorded with a θ offset of 1º; c) Histogram and normal distribution of Pt nanoparticles in the channels*

A uniform coverage of the micro-channels walls was successfully achieved *via* 80 ALD cycles. From the SEM characterization and especially analyzing the images taken with a backscattered electrons detector (figure 4.a), one can notice that a uniform coverage along the micro-channels walls was obtained. The presence of platinum was also confirmed by XRD (figure 4.b). A broad peak around 39.7º was observed, either corresponding to the (111) reflection of Pt metal (JCPDS 00-001-1190) or (020) and (200) reflection of $PtO_2$ (JCPDS 00-043-1045). A reducing treatment in 5%-$H_2$/Ar atmosphere at 300 ºC was therefore applied prior to DME reforming experiments to avoid any platinum dioxide in the micro-reactor.

The particle size distribution is reported in figure 4.c, where particles with diameter size ranging from 5 to 14 nm were observed. The average diameter size estimated is 9.6 nm. An estimation of the platinum surface coverage can be made accounting for the over 300 particles $\mu m^{-2}$ observed along the micro-channels that is around 3% of the exposed surface. Correspondingly, the metal loading can be estimated assuming hemispherical shape of the particles and it was found to be around 1%wt of the alumina support layer.

For an increasing number of ALD cycles, a broader particle distribution was indeed expected as shown before by Rontu *et al*. [41]. Surface chemical modification is thereby pointed out as an effective route to obtain more hydroxyl-terminate surface, thus reducing the number of

ALD cycles and obtaining a narrower distribution and smaller particles, usually more active in catalysis.

## 3.2. DME gas conversion

### 3.2.1 Temperature effects

The effect of temperature on DME conversion and product selectivity was investigated by placing the micro-reformer unit inside a customized fixed bed reactor and into a furnace.

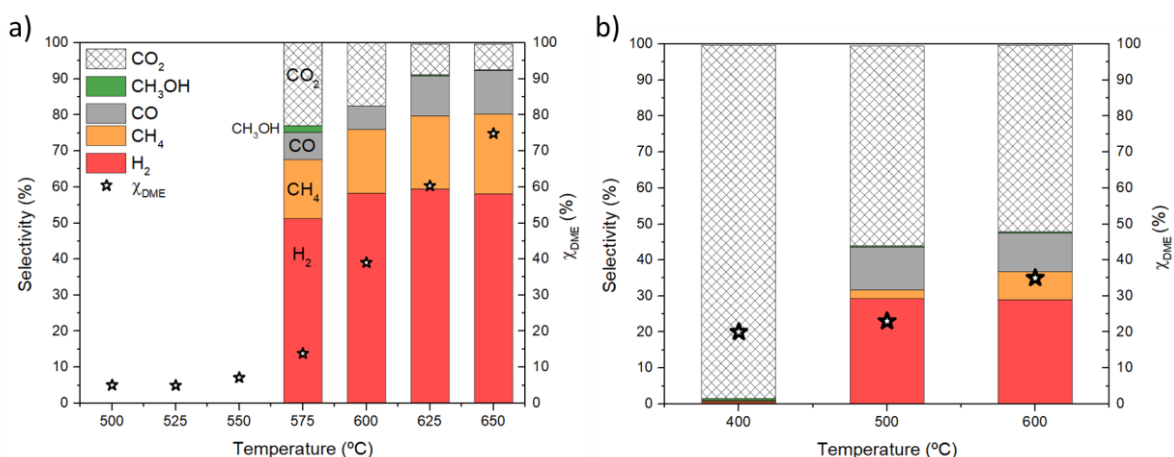

*Figure 4. Product selectivity (left Y-axis) and DME conversion (right Y-axis) as a function of temperature for a) SR reaction and b) POX reaction. The same legend is used for both graphs.*

For the steam reforming reaction, a temperature range between 500 ºC and 650 ºC was selected for the experiment. The main products of reaction are $H_2$, $CH_4$, $CH_3OH$, CO and $CO_2$. As shown in figure 4.a, the DME conversion is below 10% at 500 ºC, while further increasing up to 74% at 650 ºC. Hydrogen selectivity close to 60% is observed above 600 ºC. Selectivity of the products in only reported for temperatures above 575 ºC where the error in the calculation was found acceptable.

As expected, the DME conversion is quite low at temperature lower than 575 ºC. Methanol is observed at the lower temperature range highlighting the role of the acidic $Al_2O_3$ active support to the hydration of DME (equation 1). At higher temperatures, $CH_3OH$ disappears suggesting a complete methanol steam reforming (equation 2) is occurring. Here, selectivity to $H_2$ reaches values as high as 60%. However, as the temperature increases the amount of $CH_4$ and CO rises because of both the reverse water-gas shift reaction (r-WGS, eq. 5) and DME decomposition (eq. 6) are favored at higher temperature [42,43]:

$$CO_2 + H_2 \rightleftarrows CO + H_2O \tag{5}$$

$$CH_3OCH_3 \rightleftarrows CH_4 + CO + H_2 \tag{6}$$

This trend leading to $CH_4$ formation was already reported for Pt-loaded $Al_2O_3$ catalysts, where the addition of Pd to the catalytic system helped inhibiting methane formation [34].

The effect of temperature on DME POX was studied within a lower temperature range (i.e. 400-600 ºC), as it is expected to take place at lower temperature than steam reforming. Hydrogen selectivity of about 30% and DME conversion of 36% were obtained.

At 400 ºC no hydrogen was detected while carbon dioxide was present due to a low-temperature complete oxidation of the fuel (eq. 7).

$$CH_3OCH_3 + 3O_2 \rightleftarrows 2CO_2 + 3H_2O \tag{7}$$

The main products of DME POX reaction, carbon monoxide and hydrogen, were found at temperature greater than 460 ºC. An increasing concentration of methane is observed at higher temperature, which is associated to DME decomposition (eq. 6), since CO methanation is not favored at high temperature due to the chemical equilibrium. The $H_2$ yield in the transient regime during start-up of the device was further characterized and will be presented in the following section for the standalone device.

The hydrogen-to-carbon monoxide ratio was found to be around 2.5, higher than the stoichiometric value of 1.5. This result can be attributed to the water-gas shift reaction where CO reacts with the steam produced by DME full oxidation.

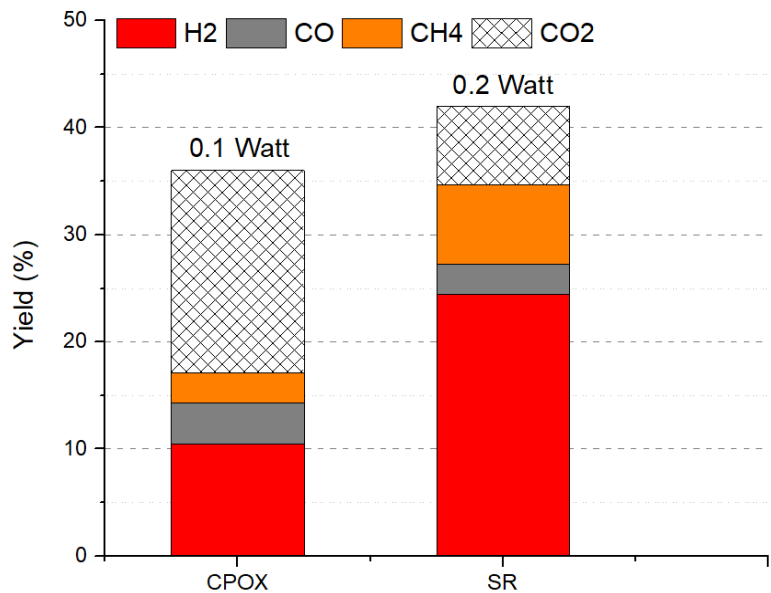

*Figure 5. Comparison between yield of products for POX and SR reactions at T=600 ºC and resulting power obtained by an ideal fuel cell power generator (eq. 4)*

When comparing the amount of hydrogen produced by these two processes at 600 °C in figure 5, it is noticeable that steam reforming leads to the generation of twice the amount of hydrogen per mole of reformed gas with respect to the POX reaction. Overall, the syngas production from DME SR doubles its POX counterpart.

The results shown above demonstrate the potential for integration of such micro-reforming unit in portable power generator, delivering approximately 0.1 and 0.2 W from DME POX and SR respectively (from eq. 4), thus pointing out the present technology as a promising route for a fully scalable production of on-board fuel micro-reformers. Moreover, one must take into account that this standalone device has been especially designed for the start-up of a µ-SOFC, while a full device also comprises a bulk reforming unit[25] that would greatly enhance the syngas productivity. Indeed, the deposition technique employed in this work can be easily applied to the fabrication of different catalytic systems, including for instance Pd, Cu, Zn, Ru, Ni metal catalysts and $ZrO_2$ or $CeO_2$ active supports, to mention some.

### 3.2.2 Standalone operation of the DME POX reactor

The integration of a micro-reformer in a micro-SOFC requires an embedded internal heating element for an efficient start-up of the device. In this case, it is obtained by fabricating a thin-film W/Au serpentine heater on top of the suspended micro-channels platform in the device (see figure 1).

DME POX was chosen as the most suitable reaction for micro-reactors to be included in SOFC portable systems since its exothermic character and easier integration (without the need of steam generation on-board). Therefore, these are the only measurements presented in this section.

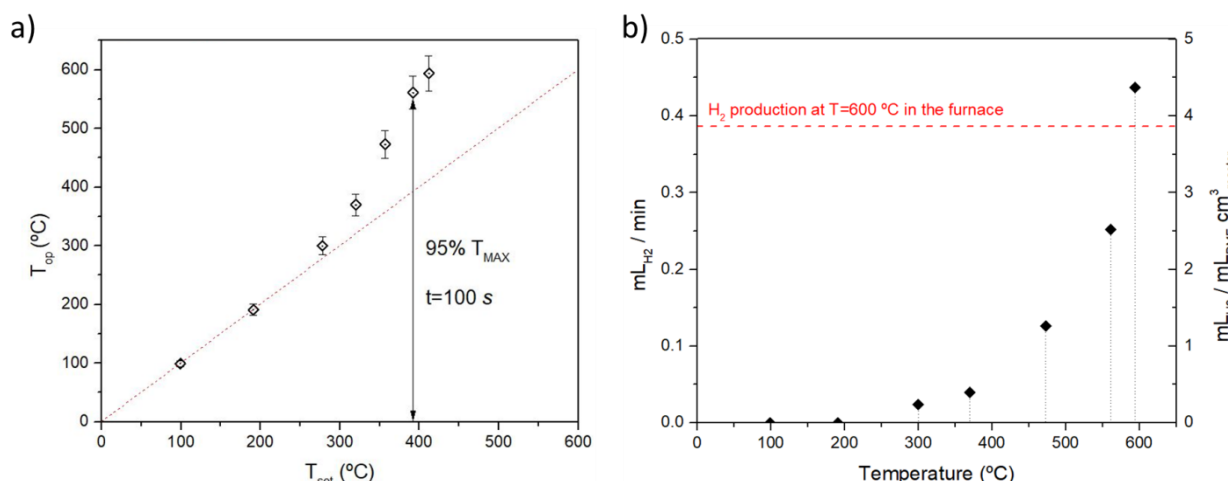

Figure 6. (a) Characteristic temperature behavior of the thin-film heater during testing with a total inlet flow rate of 15 mL $min^{-1}$; (b) hydrogen production rate (left Y-axis) and specific volumetric production of hydrogen (right Y-axis) during start-up

Figure 6.a shows the temperature evolution of the embedded heater, i.e. the reactor channels, during the start-up (when a constant power is applied and the DME:Air:$N_2$ mixture is fed into the system at a flow of 15 ml $min^{-1}$). Complementary, figure 6.b presents the hydrogen production rate generated during the same start-up transient regime. As clearly observed in the figures, the hydrogen production reaction starts around 300 ºC. Once this occurs, the difference between the temperature set point (without gas flowing) and operating temperature becomes evident, ultimately reaching a maximum temperature of 595 ºC when the temperature was set at 400 ºC. In other words, the system is self-heated by the exothermal reaction when it starts. Beyond this point, the thermally self-sustained mode is enabled. For the measurement setup developed in this work, the self-sustained mode is reached after 100 *s*, which is thrice the time reported by the authors for measurements carried out under vacuum[24]. Although the reactor embedded in the final system is expected to behave approaching the values under vacuum, the energy consumption required here for the start-up was still below 500 J, a value considered feasible for a hybrid heating solution based on electrical heating and catalytic oxidation[25]. Moreover, these results show that above 300 ºC

the hydrogen production exponentially increases due to the temperature increase, achieving values of hydrogen yield surprisingly higher than those reported in section 3.2.1 at the same temperature. Further work is on-going to fully understand this behaviour and the potential effects of local heating.

Finally, figure 6.b (right Y-axis) shows great values of specific volumetric production of hydrogen from the volume of the fuel of choice (~ 4.5 $mL_{H2}$ $mL^{-1}_{DME}$ $cm^{-3}$). This is a crucial parameter for the design of an efficient catalytic conversion micro-reactor, especially when aims portable applications. In this regard, this work reinforces previous reports by the authors where a reduction of the micro-channel diameter size from one millimeter to 2 μm resulted in three orders of magnitude of enhancement of the same parameter[23].

Overall, this section proves the possibility to operate a silicon-based micro-reactor for DME conversion with excellent hydrogen production per unit volume by reaching the thermally self-sustained mode using embedded micro-heaters.

## Conclusions

In this work, wafer-level fabrication of a suspended Si-based micro-reformer unit for DME-to-syngas conversion is presented. The micro-reactor, fabricated by means of mainstream micro-fabrication processes, comprises an array of vertically-aligned micro-channels of 50 μm in diameter for a total active area of about 6 $cm^2$, a thin-film heater for efficient reaction start-up and a suspended $Si_3N_4$ membrane for thermal insulation of the active area.

The micro-channels walls were effectively coated with a Pt nanoparticle dispersion (1%wt of the active support, 9.6 nm in diameter) supported on 100 nm-thick $Al_2O_3$ layer by means of atomic layer deposition (ALD). The system was tested for both DME steam reforming (SR) and partial oxidation (POX) at intermediate temperatures. A 74% DME conversion with hydrogen selectivity around 60% was obtained at 650 ºC from DME steam reforming, while results on partial oxidation showed 36% DME conversion and 30% hydrogen selectivity at 600 ºC. The heterogeneous catalyst chosen for this proof-of-concept showed relatively high selectivity to $CH_4$ and CO because of the side reactions involved.

Efficient POX reaction start-up (energy consumption < 500 J) was demonstrated in standalone configuration using a 3D printed ceramic holder. State-of-art specific volumetric hydrogen production rates of 4.5 $mL_{H2}$ $mL^{-1}_{DME}$ $cm^{-3}$ were achieved after reaching thermally self-sustained mode using the embedded micro-heater.

All in all, this work shows the great potential of monolithic MEMS-based micro-reactors (and their flexibility to reform different types of fuels depending on selective deposition of catalysts) for the final application in on-board hydrogen production in portable systems.

## Acknowledgement

This project has received funding from the European Research Council (ERC) under the European Union's Horizon 2020 research and innovation programme (ERC-2015-CoG, grant agreement No #681146- ULTRA-SOFC). LS is grateful to MICINN Ramon y Cajal program for an individual fellowship grant agreement RYC2019-026704-I. JL is a Serra Húnter Fellow and is grateful to ICREA Academia program and projects MICINN/FEDER

RTI2018-093996-B-C31 and GC 2017 SGR 128. Special thanks go to the IMB-CNM cleanroom staff.